\documentclass[aps,twocolumn,groupedaddress,nobalancelastpage]{revtex4}
\usepackage{graphicx}
\usepackage{dcolumn}
\usepackage{bm}
\usepackage{hyperref}
\usepackage{subfigure}

\begin{document}

\title{21 cm radiation  - a new probe of variation in the fine structure constant}
\author{Rishi Khatri}
\email{rkhatri2@uiuc.edu}
\affiliation{Department of Astronomy, University of Illinois at Urbana-Champaign, 1002 W.~Green Street, Urbana, IL 61801}
\author{Benjamin D. Wandelt}
\email{bwandelt@uiuc.edu}
\affiliation{Departments of Physics and Astronomy, University of Illinois at Urbana-Champaign, 1002 W.~Green Street, Urbana, IL 61801}

\date{\today}
\begin{abstract}
We investigate the effect of variation in the value of the fine structure constant
($\alpha$)
at high redshifts (recombination $ > z > 30$ ) on the absorption
of the 
cosmic microwave background (CMB) at 21
cm hyperfine transition of the neutral atomic hydrogen. We find that the 21
cm signal is very sensitive to the variations in $\alpha$ and it is so
far the only probe of the fine structure constant in this redshift range.
 A change in the value of $\alpha$
 by $1 \%$ changes the mean
 brightness temperature decrement of the CMB due to 21 cm absorption by $>
 5 \%$ over the redshift range $z < 50$.  There is an effect of similar magnitude on the amplitude of the
 fluctuations in the brightness temperature. The redshift of maximum
 absorption also changes by  $\sim 5\%$.
\end{abstract}
\maketitle

\noindent \emph{Introduction. ---} Why do the fundamental constants of
nature take the values that we measure ? This question was
 posed as a fundamental problem of physics in the last century. Dirac
\cite{dirac1937,dirac1938} first considered the question of variation of
these constants with time. The standard model does not explain 
the values of these constants, especially the constants
determining the strength of the four fundamental forces. These constants can
 indeed vary naturally, though not necessarily, in space as well as time in  
Grand Unified Theories (GUTs) and theories of quantum gravity \cite{uzan}.  Thus a measurement of variation or a constraint on
non-variation of the fundamental constants is an important probe of new
physics beyond the standard model and general relativity.
Testing the variation of 
constants has become more important in light of  current data indicating
the presence of dark energy. Dark energy could be a cosmological constant,
which fits current data \cite{wmap3}, but could also be
evidence for physics beyond the
standard model. It is therefore important to find new ways to distinguish
different models of dark energy. If dark energy couples to the standard particle physics
 it could cause variations in the fundamental constants of the standard
model \cite{caroll}. 
Testing for variations in the fundamental constants thus
also probes the properties of dark energy \cite{chiba,parkinson} and hence
is important for any attempt to explain dark energy theoretically.

In this  \emph{Letter} we propose a new probe of variations in the fine
structure constant on cosmological time scales. The most stringent existing
constraints on the value of the fine structure constant in the early
Universe are from the measurements
of  quasar spectra involving fine structure transitions. These measurement suggest a change
$|(\alpha_t - \alpha)/\alpha| =  |\delta\alpha/\alpha| \lesssim  10^{-5}$, where $\alpha_t$
is the value of $\alpha$ at time t \cite{webb,murphy,chand}. For these quasar measurements t corresponds to a
 $z < 3.5$.  If
 the variation is monotonic $|\delta\alpha|$ would be larger at higher
 redshifts. If the variation is not monotonic, then measurements at many
 different redshifts would be required to trace the features in its evolution. The constraints at very early times, $z > 10$, are not very
stringent and come from the CMB ($z \sim 10^3$) and big bang nucleosyntheis
(BBN, $z \sim 10^9-10^{10}$).  From CMB
\cite{ichikawa,rocha,avelino} $|\delta\alpha/\alpha| < 3-9 \%$ and from BBN \cite{cyburt} $|\delta\alpha/\alpha|
< 6\%$.  There are no constraints for redshifts in the range $1000 > z >
10$. 

The 21 cm absorption of CMB provides an opportunity to
constrain $\alpha$ during these ``dark ages.'' Also the 21 cm absorption signal is
available for a range of redshifts during this period and thus can be a
useful probe for tracing the evolution of $\alpha$. One advantage of this method
is that we are measuring the  amount of absorption of the CMB radiation which depends on the
Einstein emission/absorption coefficients. As we will see below the
Einstein coefficients are more sensitive to changes in $\alpha$
($A_{10}\propto \alpha^{13}$) than the fine structure/hyperfine structure
splitting itself.

\noindent \emph{21 cm radiation from the dark ages. ---} The 21 cm signal
from neutral hydrogen  after recombination and before
re-ionization has been investigated by many authors
\cite{hogan,scott,loeb,bhardwaj}. We refer to \cite{review} for detailed
review. After recombination the radiation temperature goes
down as $1 + z$. The baryons however are prevented from cooling
adiabatically due to the small amount of residual electrons which couple
the gas to the radiation through Thomson scattering. At $z \sim
 200$ this process becomes inefficient and the matter decouples thermally from
the radiation. The hydrogen atoms after recombination are in the ground state which
is split into a singlet and a triplet state due to hyperfine splitting.
 The occupation of the excited triplet state and the
lower energy singlet state can be described by defining the spin
temperature by the relation 
$n_t/n_s = g_t/g_se^{-T_{\star}/T_s}$,
 where $n_t$ and $n_s$ are the number densities of atoms in triplet
 and
 singlet states respectively, $g_t$ and $g_s$ are the corresponding
 statistical weights with $g_t/g_s = 3$, $T_{\star} = 0.068 K$ is the energy
 difference between the two states and equals 21 cm in wavelength units and $T_s$
 is the spin temperature \cite{field}.

The evolution equations for the gas temperature $T_g$, radiation
temperature $T_{\gamma} = 2.726(1+z) \rm{K}$,
ionization fraction $x = n_e/n_H$, where $n_e$ is the number density of electrons and $n_H$ is the total number density of hydrogen nuclei, and spin temperature $T_s$ can be written as \cite{ma,peebles,bhardwaj}
\begin{equation}
\label{ts}\frac{dT_s}{dz}=\frac{4T_s^2}{H\left(1+z\right)}\left(\frac{1}{T_g}-\frac{1}{T_s}\right)C_{10}+\left(\frac{1}{T_{\gamma}}-\frac{1}{T_s}
\right)T_{\gamma}\frac{A_{10}}{T_{\star}}\end{equation}
\begin{equation}
\label{tg}\frac{dT_g}{dz}=\frac{2T_g}{1+z}-\frac{8\sigma_T}{3m_ec}\frac{4\sigma_{SB}}{c}\frac{x}{1+x_{He}+x}\frac{T_{\gamma}^4}{H\left(1+z\right)}\left(T_{\gamma}-T_g\right)\end{equation}
\begin{equation}
\label{ion} \frac{dx}{dz}=\frac{-C_r}{H\left(1+z\right)}\left(\beta\left(1-x\right)-n_H\alpha_2x^2\right),
\end{equation}
where H is the Hubble parameter at redshift z, $C_{10}=\left(\kappa_{10}^{HH}n_H
+\kappa_{10}^{eH} x n_H \right)$,  $\kappa_{10}^{HH}$ is the collisional
de-excitation rate from triplet to singlet state for H-H collisions
\cite{allison,zyg05}, $\kappa_{10}^{eH}$ is the corresponding cross section
for e-H collisions \cite{furlanetto2}, $A_{10}$ is
the Einstein A coefficient for spontaneous transition, $\sigma_T$ is the Thomson
scattering cross section, $\sigma_{SB}$ is the Stephan-Boltzmann constant,
$m_e$ is the mass of electron and c is the speed of light in vacuum
. $x_{He} = n_{He}/n_H$, $n_{He}$ being the number density of Helium
nuclei. $C_r$, $\beta$ and $\alpha_2$ are defined by
the equations \cite{ma,peebles,recfast}
\begin{eqnarray}
\alpha_2 & = &
F
10^{-13}\frac{4.309(T_g/10^4)^{-0.6166}}{1+0.6703(T_g/10^4)^{0.5300}}\hspace{2
  pt}\rm{cm^3}\hspace{2 pt}\rm{s^{-1}}
\\
\beta & = & \left(\frac{m_ek_BT_g}{2\pi\hbar^2}\right)^{3/2}e^{-B_1/k_BT_g}\alpha_2
\\
C_r & = & \frac{\Lambda_{\alpha} + \Lambda_{2s\rightarrow
    1s}(1-x)n_H}{\Lambda_{\alpha} + (1-x)n_H(\Lambda_{2s\rightarrow 1s} + \beta e^{hc/k_BT_g\lambda_{\alpha}})}
\end{eqnarray}
where $\Lambda_{\alpha} = 8\pi H
  \left(1+z\right)/\lambda_{\alpha}^3$, $\lambda_{\alpha} = 8 \pi \hbar
  c/3B_1$ and $B_1=\alpha^2m_ec^2/2$. $F=1.14$ is the fudge factor to take
  into account the non-equilibrium among higher energy levels of hydrogen
  \cite{recfast}. Helium recombination can be ignored since it has no
  effect on the 21 cm signal.
Solving equations (\ref{ts}-\ref{ion}) in a given cosmology gives the evolution of spin
temperature with redshift. The change in the brightness temperature of the
CMB at the corresponding wavelength is then given by \cite{field,field2}
\begin{eqnarray}
\label{tb}T_b = \frac{\left(T_s - T_{\gamma}\right)\tau}{\left(1+z\right)},
\hspace{10 pt}
\tau=\frac{3c^3\hbar A_{10}n_H}{16k_B\nu_{21}^2H T_s},
\end{eqnarray}
where $\nu_{21} = k_BT_{\star}/h \sim 1420 MHz$. The brightness temperature
is related to the observed intensity by the Rayleigh-Jeans formula $T_b =
I_{\nu}c^2/2k_B\nu^2$, where $I_{\nu}$ is the specific intensity and $\nu$
is the frequency of observation.

There will also be fluctuations in $T_s$ and $T_b$ due to inhomogeneities in $n_H$ and
$T_g$ which are related to the primordial inhomogeneities in the
gravitational potential \cite{loeb,bhardwaj}. The angular power spectrum is
then given by $C_l(z) = \langle a_{lm}a_{lm}^*\rangle$, where $a_{lm}$ are the expansion
coefficients in the spherical harmonic expansion of $\delta T_b =
T_b-\bar{T_b}$, where $\bar{T_b}$ is the mean brightness temperature.
We follow \cite{bhardwaj} in our calculations. The baryon power spectrum is
calculated using CMBFAST \cite{cmbfast}.

\noindent \emph{Effect of change in $\alpha$. ---} A different $\alpha$ during recombination affects the CMB power spectrum
due to Thomson scattering of photons through equation (\ref{ion}) and the
Thomson scattering cross section \cite{kap,han}. As with CMB we ignore the
variation in the fudge factor, $F$, with $\alpha$ in Eq. (\ref{ion}) since its effect is
negligible. The crucial point is that
for the 21 cm transition,
there is the following additional dependence on $\alpha$ 
in equations (\ref{ts}), (\ref{tg}) and (\ref{tb}). The Einstein A Coefficient is given by
 $A_{10}=64\pi^4\nu_{21}^3S_{21}/3hc^3g_2$, where $g_2$ is the
statistical weight of excited state, $S_{21} = 3\beta_M^2$, $\beta_M=
eh/4\pi m_ec$ being the Bohr magneton \cite{wild}.
Now $\nu_{21}\propto \alpha^2R_{\infty} \propto \alpha^4$ and $T_{\star}
\propto \nu_{21}$, where $R_{\infty}$ is
the Rydberg's constant. 
Thus we see that the spontaneous emission coefficient, $A_{10}\propto
\nu_{21}^3\beta_M^2 \propto \alpha^{13}$, is a very sensitive
function of $\alpha$.

The $\alpha$ dependence of the collisional de-excitation rate is more
complicated. We use  ab-initio calculations and asymptotic formulae for
large separations 
 of  potential energy
curves of  the ground state and the first
excited triplet state of hydrogen molecule to calculate the spin change
collision cross sections \cite{dalgarno,herring,frye,kolosrych90,wol93,wol95}. These are nothing but the total energy at a given separation in
the clamped nuclei approximation or 
the expectation values of the electronic Hamiltonian, $H_e$, which has a kinetic
energy term ($T$) and a Coulomb potential term ($V$):
\begin{equation}
H_e = T + V
\end{equation}
A change in the fine structure constant can be treated as a perturbation in
the Coulomb potential ($V\propto \alpha$). Therefore if $\delta$ is the
fractional change in alpha, so that
$\alpha_{new}=\alpha\left(1+\delta\right)$, then $V_{new} =
V\left(1+\delta\right)$. Now to first order in perturbation theory the
expectation value of the new Hamiltonian is given by 
\begin{equation}
\langle H_{new} \rangle = \langle T + V_{new} \rangle = \langle H_e \rangle
+ \delta \langle V \rangle.
\end{equation}
In above the $\langle \rangle$ denote the expectation value over the unperturbed
wavefunction everywhere.
Also the change in the derivative to first order is given by $\delta dV/dR$. We
estimate $dV/dR$ by fitting a polynomial function to the V-R curve from the
ab-initio calculations \cite{dalgarno,herring,frye,kolosrych90,wol93,wol95}. Thus from the
ab initio potential energy curve  at unperturbed $\alpha$, we can construct the first order
corrections when $\alpha$ changes by a small amount. We first calculate the scattering phase shifts by
integrating the partial wave equations and use them to calculate the scattering
cross sections and the rate coefficients ($\kappa_{10}^{HH}$) using the standard
scattering theory \cite{mott,dalgarno2,allison,zyg05}. 

\begin{figure}
\includegraphics{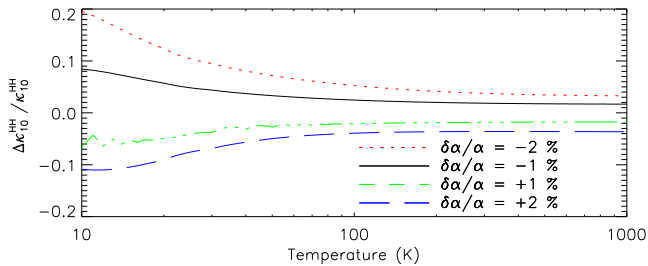}
\caption{\label{cross} The fractional change in $\kappa_{10}^{HH}$ with respect to the unperturbed value.}
\end{figure}

We checked our code by comparing our cross sections for unperturbed $\alpha$
with those calculated by Zygelman and Allison $\&$ Dalgarno
\cite{allison,zyg05}. They agree to better than $1\%$ for $T_g > 40K$ and
to $2\%$ for $15K < T_g < 40K$. The small error that we make at low temperatures
is due to our ignoring the higher order effects which have been taken into
account in the calculations in \cite{zyg05}. We verified that this small disagreement
at
low temperatures has a negligible effect on our results by repeating all
the calculations using the cross sections in \cite{allison,zyg05} for
unperturbed $\alpha$. Fig. \ref{cross} shows $\kappa_{10}^{HH}$ and
the fractional change in $\kappa_{10}^{HH}$ as a 
function of temperature for different values of 
$\alpha$. Changes in $\kappa_{10}^{eH}$ due to variation in $\alpha$ has
negligible effect on $T_b$ and can be ignored.

\begin{figure}
\includegraphics{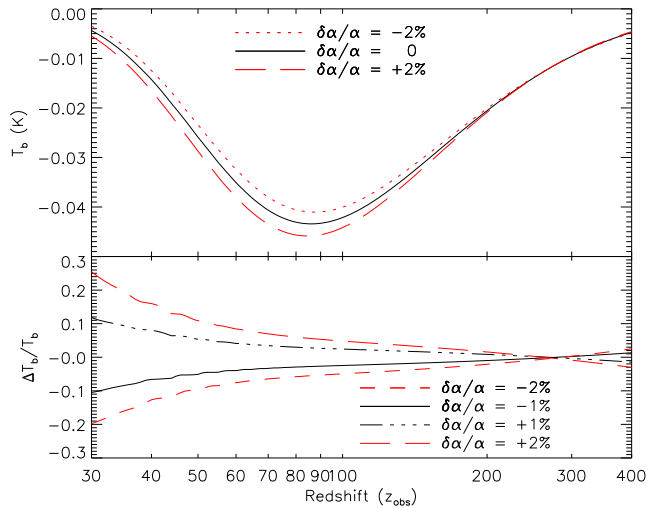}
\caption{\label{bright} Upper panel shows $T_b$ with  different
values of $\alpha$. Bottom panel shows the fractional change $(T_b\left(\alpha\right) - T_b)/T_b$.}
\end{figure}

\begin{figure}
\includegraphics{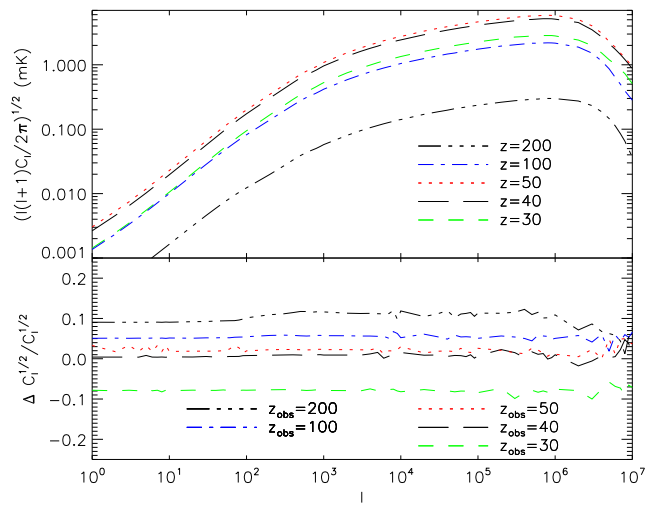}
\caption{\label{ps} Upper panel shows the angular power spectrum
  $\sqrt{l(l+1)C_l/2\pi}$ at several redshifts. Bottom panel shows
  $(\sqrt{C_l\left(\delta \alpha=-2\%\right)}-\sqrt{C_l})/\sqrt{C_l}$ for
  the same redshifts.}
\end{figure}

\noindent \emph{Results and Discussion. ---} We use $\Lambda CDM$ cosmology with WMAP3 parameters \cite{wmap3} in our calculations.
 Fig. \ref{bright} shows the observable $T_b$ as a function of
observed redshift ($z_{obs}\equiv \nu_{21}^{now}/\nu_{obs} - 1$) where $\nu_{obs}$
is the observed frequency today.  From
Eq. (\ref{tb}) $T_b \propto A_{10}/\nu_{21}^2 \propto \alpha^5$ giving $\Delta
T_b/T_b = 5\%$ for $1\%$ change in $\alpha$.  This is approximately the
 change we see in fig. \ref{bright}. Change in $\alpha$ also changes
the coupling of $T_s$  to $T_{\gamma}$  which is
opposite to the above mentioned $\alpha^5$ effect. Also $z_{obs}$
corresponds to different $z$ for different values of $\alpha$ causing
additional change in the signal.
It is clear that the 21 cm brightness
temperature is a sensitive probe of the variations in the fine structure
constant. The maximum relative change in the brightness temperature ($\Delta
T_b/T_b$) is $> 5\%$ at $z_{obs} < 50$  for a $1\%$ variation in
$\alpha$. 

Variations in $\alpha$ also affects the 
power spectrum of the spatial fluctuations. This is shown in
fig. \ref{ps} for $\delta \alpha = -2\%$. The amplitude of the fluctuations is proportional to
$T_b$. Thus we expect the
amplitude of the power
spectrum to have similar dependence on $\alpha$ as $T_b$. There is however
also contribution due to fluctuations in $T_s$ due to inhomogeneities in
$n_H$ and $T_g$. This additional effect causes an increase in the sensitivity
to variations in $\alpha$ compared to  $T_b$ at $z_{obs}>100$ and a decrease in
sensitivity at $z_{obs}<100$. There is also a change in the sign of $\Delta
\sqrt{C_l}$ at $z_{obs}\sim 40$ which is different from $T_b$, where this occurs at $z_{obs}\sim 280$. This change in sign is characteristic of the
$\alpha$ dependence.

The detectability of the signal is limited by noise due to foregrounds. The
noise  can be expressed
in temperature units for a single aperture telescope \cite{kraus} as $T_N = T_{sys}/\epsilon \left(\Delta \nu
  t_{int}\right)^{1/2}$
where $\epsilon$ is
the aperture efficiency which is close to unity, $\Delta \nu$ is the bandwidth, $t_{int}$ is the
integration time and $T_{sys} $ is the system temperature which at low
frequencies is the temperature of the galactic foregrounds $\sim
19000\rm{K}$ at 22 MHz in a quiet portion of sky \cite{22}. $T_N \sim 1.4 \rm{mK}$
for $\Delta \nu\sim 4MHz$ and $t_{int}=2000hrs$.  This means that the
sensitivity of a single station of a telescope like LWA \cite{lwa} or LOFAR
\cite{lofar}
can give a constraint on $\Delta \alpha$ of $\sim 0.85\%$, improving to
$\sim 0.3\%$ for the full LWA. The fundamental challenge to realizing this
measurement is the required precision to which foregrounds have to be
subtracted. 

 Although the foreground removal from the
mean signal may prove to be impossible, this problem may be overcome by
using the effect of
$\delta \alpha$ on the angular
power spectrum of fluctuations in the 21cm absorption. Several promising foreground removal
strategies have been proposed in the literature
\cite{santos,morales,wang,zald}. To obtain a similar sub-percent constraint
from the angular power spectrum requires the higher sensitivity at small
angular scales ($T_N \sim
0.1\rm{mK}$) of a larger telescope (for example a low frequency
equivalent of SKA ,http://www.skatelescope.org)  or much longer
observation times \cite{loeb}. This is likely to be feasible
with future technological advancement.

This new probe is complementary to CMB
and BBN since it is in a different redshift range and has the potential to
provide constraints comparable to the CMB experiment Planck \cite{rocha}.
The 21 cm signal is of course also sensitive to the cosmological parameters. A $1\%$
change in the baryon density, $\Omega_b$, has a $\sim 2\%$ effect on $T_b$
while a $1\%$ change in the Hubble parameter changes $T_b$ by $\sim
3\%$. Similar change in the Helium fraction from BBN and the matter density, $\Omega_m$, change $T_b$ by  $<
0.5\%$. Future CMB\cite{planck} and large scale structure 
experiments \cite{des,lsst} will be able to determine these parameters to
$\sim 1\%$. As seen earlier, the variation in the 21 cm signal due to
 variation in $\alpha$ shows a characteristic dependence on $z$. This is difficult
to mimic by changing the cosmological parameters and provide a way to
 disentangle the two. In the
power spectrum there will be additional degeneracy due to
the initial conditions which might be difficult to separate. The 21 cm
signal is also affected by
variations in the gravitational constant (G) and the electron to proton mass
ratio ($\mu$). A complete treatment should consider
variations in all the constants simultaneously. 

We note that the 21 cm observations can
also provide a test for the spatial variations of $\alpha$. This can
then be used, in principle, to probe the spatial perturbations in the dark energy
in the framework of theories of quantum gravity which relate $\alpha$ to
dark energy.

\noindent \emph{Acknowledgments. ---} We are grateful to Prof. Ben McCall, Prof. Todd Martinez and Prof. Bernard
Zygelman for help in understanding the quantum chemistry calculations of
the collision cross sections. We also thank Prof. Ed Sutton and Antony
Lewis for helpful
discussions and comments. This work was partially supported by University of Illinois.
 
\bibliography{kw2007_2}

\appendix*
\section{}
Plots corresponding to figures 2 and 3 but plotted and compared at  actual
redshift $z$. They are also available as EPAPS Document
No. E-PRLTAO-98-004710,  http://www.aip.org/pubservs/epaps.html.
\begin{figure}[h]
\includegraphics{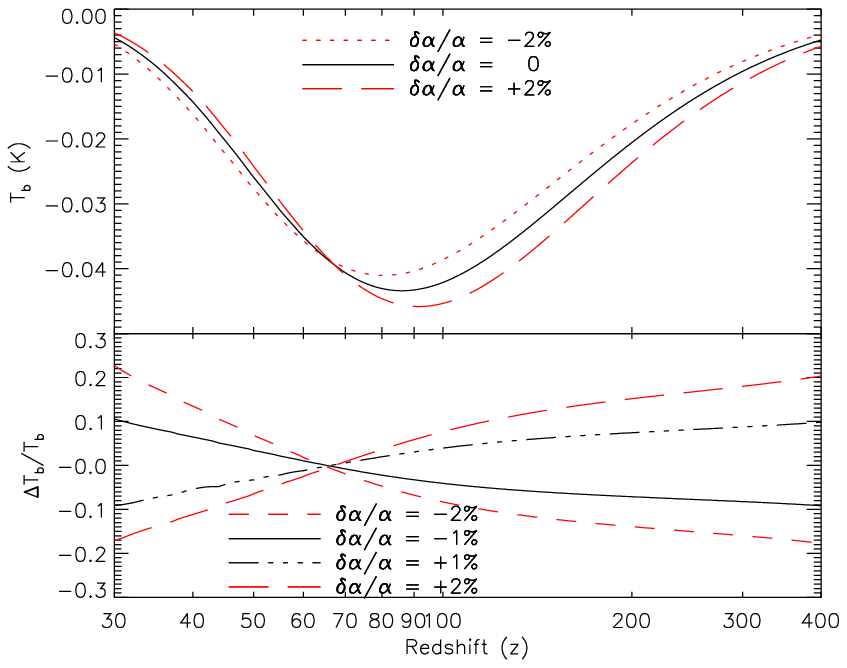}
\caption{\label{bright} Upper panel shows $T_b$ with  different
values of $\alpha$ at actual redshift z. Bottom panel shows the fractional
 change $(T_b\left(\alpha\right) - T_b)/T_b$.}
\end{figure}

\begin{figure}[h]
\includegraphics{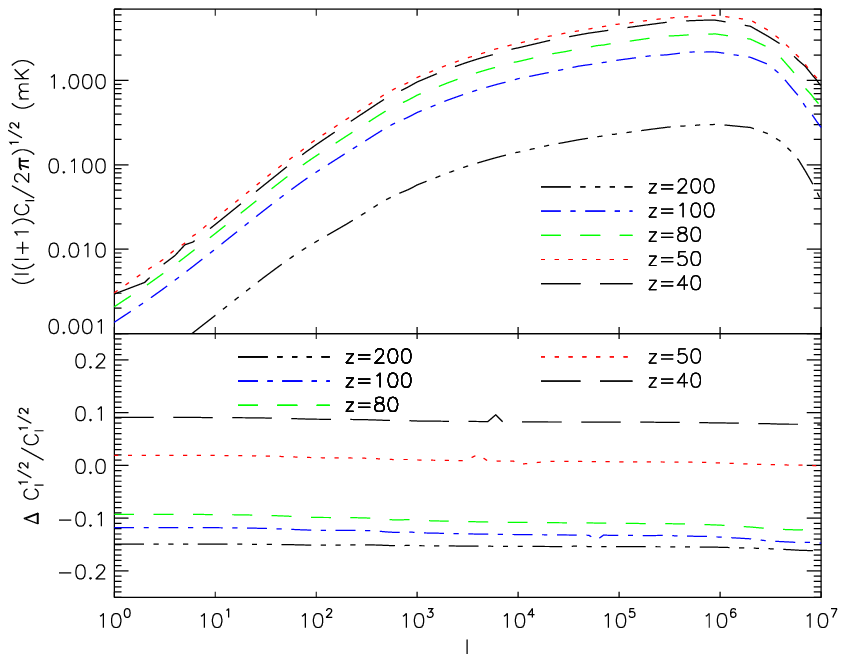}
\caption{\label{ps} Upper panel shows the angular power spectrum
  $\sqrt{l(l+1)C_l/2\pi}$ at several redshifts. Bottom panel shows
  $(\sqrt{C_l\left(\delta \alpha=-2\%\right)}-\sqrt{C_l})/\sqrt{C_l}$ for
  the same redshifts, each  compared at actual redshift z.}
\end{figure}

\end{document}